\documentclass[12pt,preprint,natbib]{aastex}
\input epsf
\def\s2n{S^{\prime}/N}

\def\astroph{0}

\begin{document}
\title{Structure Function Scaling of a 2MASS Extinction Map of Taurus.}

\author{Paolo Padoan\footnote{padoan@jpl.nasa.gov},
Laurent Cambresy and William Langer}
\affil{Jet Propulsion Laboratory, 4800 Oak Grove Drive, MS 169-506,
     California Institute of Technology, Pasadena, CA 91109-8099, USA}

\begin{abstract}

We compute the structure function scaling of a 2MASS extinction map 
of the Taurus molecular cloud complex. The scaling exponents
of the structure functions of the extinction map follow the Boldyrev's
velocity structure function scaling of super--sonic turbulence. This 
confirms our previous result based on a spectral map of $^{13}$CO
J=1-0 covering the same region and suggests that supersonic turbulence 
is important in the fragmentation of this star--forming cloud. 

\end{abstract}

\keywords{
turbulence -- ISM: dust, extinction -- ISM: kinematics and dynamics   
}

\section{Introduction}

Stars are formed predominantly from very large clouds of cold interstellar
gas containing up to millions of solar masses of material. 
The dynamics of such clouds is therefore a crucial ingredient in the process of
star formation. Observations of emission spectra of molecular transitions
have shown that the kinematics of star--forming clouds is best described
as supersonic random motions, often referred to as supersonic turbulence.
Numerical simulations of supersonic turbulence have indeed been compared
successfully with the observations, in the sense that many statistical
properties of numerical supersonic turbulent flows are also found in the
observational data (e.g. Padoan et al. 1998, 1999, 2001). 

\nocite{Padoan+98cat} \nocite{Padoan+99per} \nocite{Padoan+2001cores}

The cold gas in star--forming clouds, especially their molecular component, 
cools very rapidly down to a typical equilibrium temperature
of approximately 10~K. The shocks caused by the observed random supersonic 
velocity field are therefore roughly isothermal, which allows them 
to compress the gas effectively. Expansions are also favored by the 
isothermal behavior of the gas and large voids of very low density
can be generated. The result is a very large contrast between
the highest and the lowest densities, as commonly found in numerical
simulations with an isothermal equation of state. We usually refer to 
this effect of the turbulent velocity on the density field as 
{\it turbulent fragmentation}, to stress the fact that star--forming
clouds are more likely fragmented into dense prestellar cores directly
by the turbulent velocity field, rather than by a hierarchical process 
of gravitational fragmentation.   

The traditional way to study the dynamics of star forming clouds is
to probe their kinematics by the Doppler shift in spectral lines
of molecular transitions. However, the dynamics can also be studied
through the density field, as the gas density is so strongly affected
by the supersonic velocity. The investigation of the cloud spatial structure 
may be used to test predictions of numerical and analytical models more 
directly than using the velocity field. Projected density is in fact 
easier to measure than the velocity field. This is especially true
if stellar extinction measurements are available, since they provide 
the most reliable estimate of column density. 

Thanks to the recently completed ``Two Micron All Sky Survey'' 
(2MASS; Cutri et al. 2001),
it is now possible to generate extinction maps of several extended giant 
molecular cloud complexes with a dynamical range in both column density 
and spatial resolution that is not matched by any molecular line survey.
In this work we derive new extinction maps of the Taurus region
using 2MASS point source data, and show that we can probe values of dust 
column 
density over more than two orders of magnitude and achieve a spatial
resolution higher than in IRAS 100~$\mu$m images (\S~2). In \S~3 we 
present the results of the structure function analysis of the extinction
map and obtain a scaling that is indistinguishable from that of the 
velocity structure functions in supersonic turbulence. Conclusions
are drawn in \S~4.

\section{Extinction Maps from 2MASS Data}

Lada et al. (1994) use the stellar
extinction determined from the IR color excess, instead of stellar 
counts, to map interstellar clouds. Their method is based on superimposing 
a regular grid on the observed region, and giving each grid cell 
a value of extinction equal to the average extinction of the stars
within that cell. The number of stars per cell decreases with
increasing average extinction in the cell, because only the brightest
background stars can be seen through a large column of dust. 
The method has been improved by Lombardi \& Alves (2001), using 
Gaussian filtering to obtain a smooth regularly sampled map.

\nocite{Lada+94} \nocite{Lombardi+Alves2001}

In this work we compute stellar extinction maps with the  
method proposed by Cambresy et al. (2002). This method uses 
adaptive cells that contain a fixed number of stars instead of cells 
of fixed size. In this way it is possible to keep the spatial resolution 
as high as allowed by the local stellar density. The spatial resolution 
is higher in regions of low extinction than in regions of large extinction, 
where fewer background stars are detected and larger cells must be used. 
The average spatial resolution over the whole map can be changed
by changing the number of stars per cell. 

\ifnum\astroph=1

\begin{table}[h]
\begin{tabular}{cccc}
\hline
\hline
No. of stars & $\sigma_{A_V}$ & Median Resolution & $A_{V,{\rm max}}$  \\
             &  (mag)         & (arcmin)          & (mag) \\
\hline
100 & 0.26 & 12.4 & 8.0 \\
30  & 0.34 & 6.7  & 11.7 \\
10  & 0.49 & 3.4  & 19.5 \\
3   & 0.85 & 1.7  & 26.3 \\
1   & 1.30 & 1.0  & 32.7 \\
\hline
\end{tabular}
\caption{Extinction map parameters for different spatial resolutions.}
\label{t1}
\end{table}

\fi

The color excess is computed using the relation
$E_{H-K_s} = (H-K_s)_{\rm obs} - (H-K_s)_{\rm int}$,
where $(H-K_s)_{\rm obs}$ is the observed median color in a cell and
$(H-K_s)_{\rm int}$ is the intrinsic median color, estimated 
from the colors of supposedly unreddened stars.
In the method by Lada et al. (1994) the mean color is used instead of
the median color. We prefer to use the median color because it has 
the advantage of minimizing the effect of foreground stars, as 
shown by Cambresy et al. (2002). Visual extinction values are 
obtained from the color excess using the Rieke and Lebofsky (1985) 
extinction law, which results in the relation 
$A_V  = 15.87 \times E_{H-K_s}\label{col2av}$.

\nocite{Cambresy+2002} \nocite{Rieke+Lebofsky85}

Cambresy et al. (2002) have applied this method to study the North
America and Pelican Nebulae. However, these nebulae are
rather close to the galactic plane, and are therefore very difficult
to study with extinction measurements due to the mixture of stars and 
other clouds in their background. The Taurus molecular cloud complex
is far from the galactic plane (approximately $15^\circ$ south)
and very close to us, at a distance of approximately 140 pc 
(Kenyon, Dobrzycka \& Hartmann 1994). Contamination from foreground 
stars is therefore negligible for this region, as well as confusion 
with other clouds in the distant background.  

We have computed extinction maps of the Taurus molecular cloud complex
using 1, 3, 10, 30 and 100 stars per cell covering a region of 
$12^\circ\times10^\circ$. Figure~\ref{fig1} shows the map obtained
with 10 stars per cell. Approximately 115 known young embedded stars
(Herbig \& Bell 1988; Leinert et al. 1993; Kenyon et al. 1994; 
Briceno et al. 1998) have been excluded from the 2MASS catalog before
computing the extinction maps. The intrinsic color is computed as
the median color of stars in regions where no $^{12}CO$ (Dame et al. 
2001) is detected, within a field larger than the actual extinction 
map ($155^\circ< l <177^\circ$) and is found to be 
$(H-K_s)_{\rm int}=0.13$~mag. The standard deviation of the color
of these unreddened stars provides an estimate of the uncertainty 
(1-$\sigma$ noise) in the extinction maps.  

The value of this uncertainty, the median angular resolution (cell 
diameter) and the maximum extinction value for each map are given in 
Table~\ref{t1}. Since cell sizes are adapted to contain a fixed number 
of stars, the statistical uncertainty is independent of the extinction 
level (the noise is uniform over the map). The highest resolution achieved
with these maps is remarkable. The map with 3 stars per cell, for example, 
yields a median resolution of 1.7', almost 3 times better than the 
resolution of IRAS 100~$\mu$m images (4'$\times$5'). 
 
\nocite{Kenyon+94} \nocite{Herbig+Bell88} \nocite{Leinert+93}
\nocite{Briceno+98}

\ifnum\astroph=1


\begin{figure}[h]
\epsfxsize=8cm \epsfbox{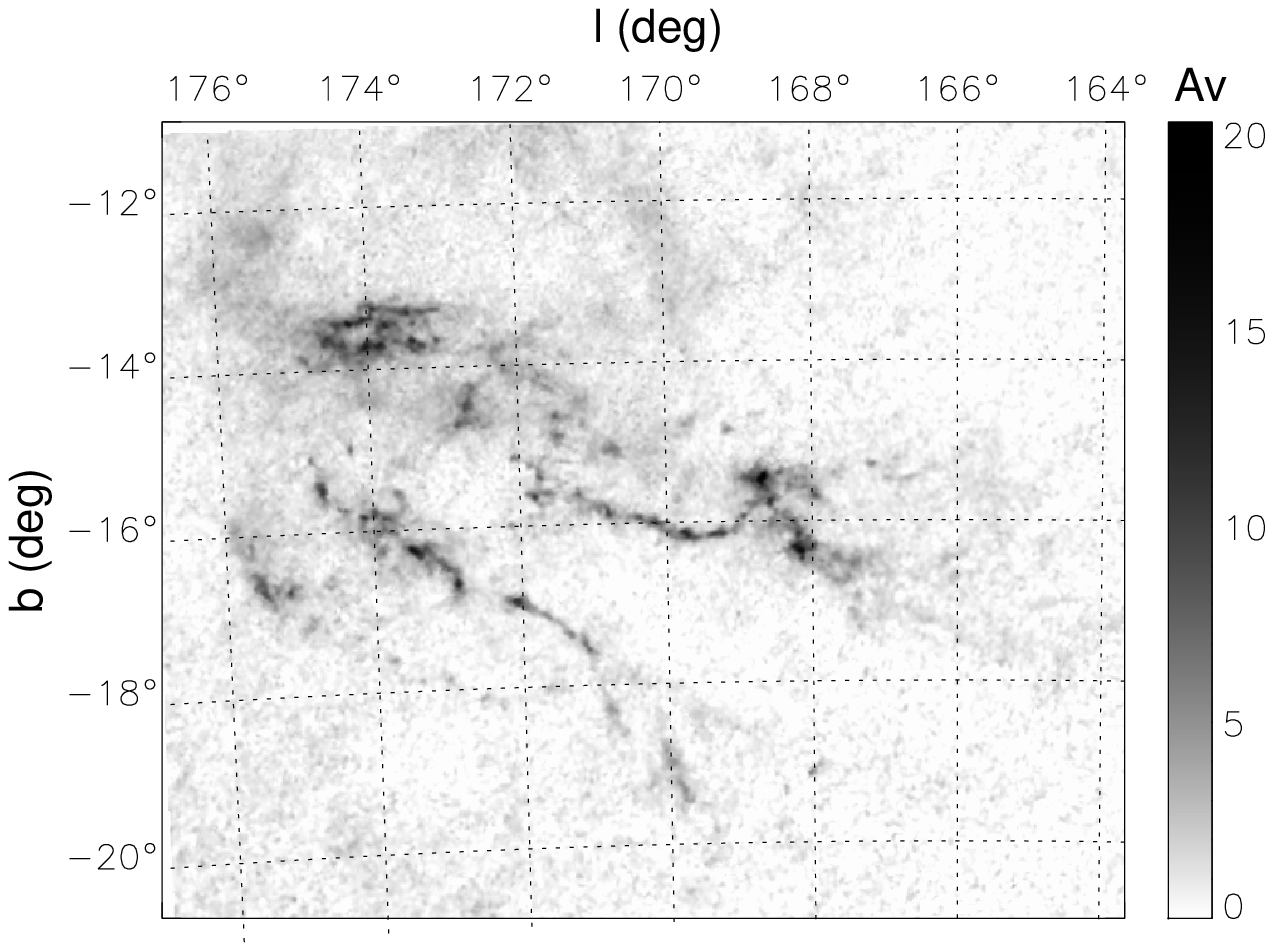}
\caption[]{Extinction map of the Taurus region computed with 10 stars
per cell.}
\label{fig1}
\end{figure}


\fi

As shown in Table~\ref{t1}, we can probe values of $A_V$ ranging 
from 0.3~mag (1$\sigma$ detection in the 100 stars per cell map) 
to 33~mag (largest extinction in the 1 star per cell map). 
Using a standard gas to dust ratio, 
$N(H+H_2)/A_{\rm V}=2\times 10^{21}$~cm$^{-2}$mag$^{-1}$ (Bohlin et al. 1978),
the range in extinction corresponds to approximately two orders of
magnitude in column density,
from $N(H+H_2)=6\times 10^{20}$~cm$^{-2}$ to 
     $N(H+H_2)=6.6\times 10^{22}$~cm$^{-2}$.

\section{Structure Functions of Projected Density}

\ifnum\astroph=1

\begin{figure}[h]
\epsfxsize=7cm \epsfbox{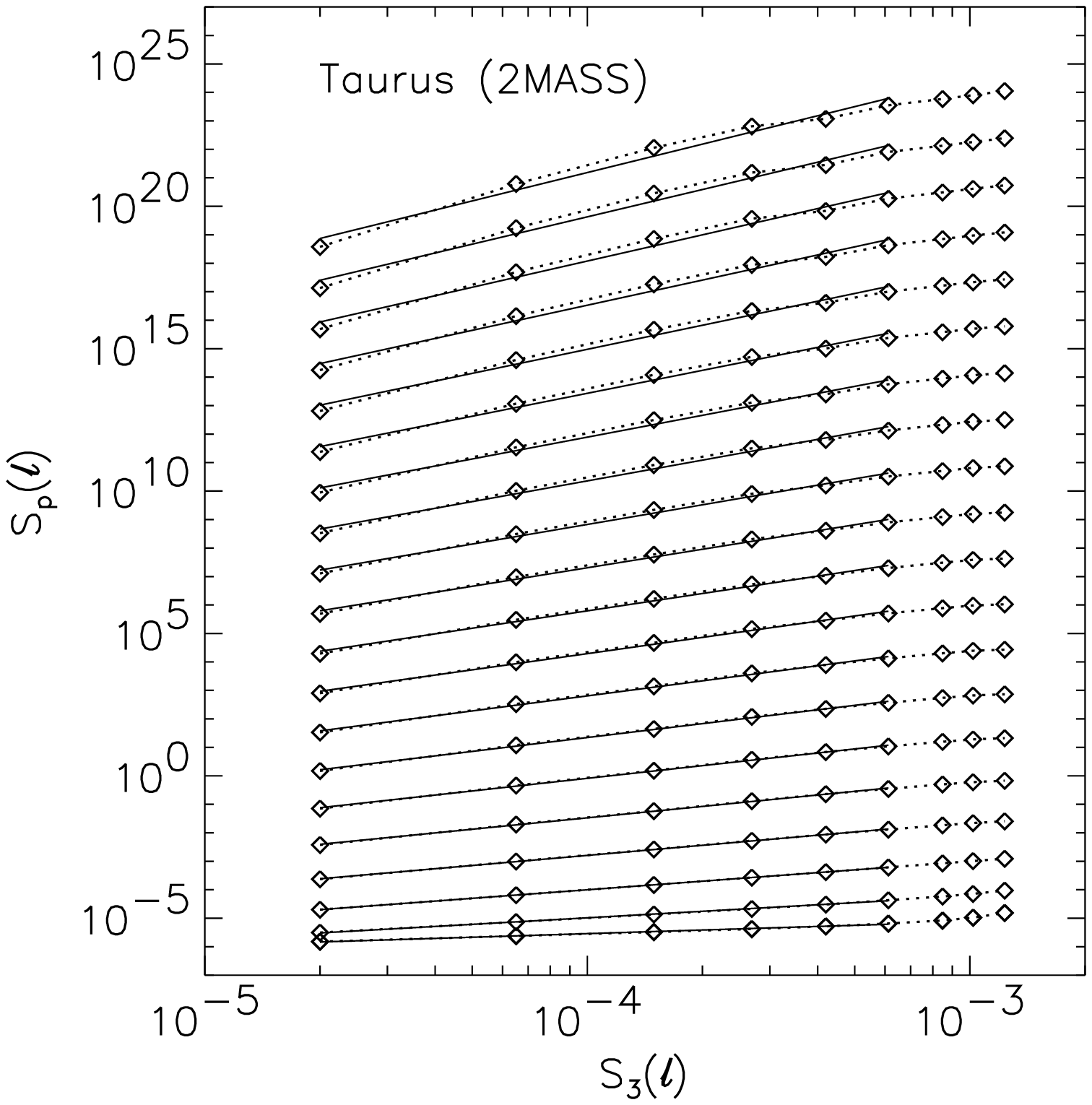}
\caption[]{Structure functions of the extinction map from $p=1$ to 20 
relative to the third order structure function. The extinction map with
10 stars per cell has been used. The solid lines show the least square
fits used to define the power law slopes.}
\label{fig2}
\end{figure}

\fi

The structure functions of the extinction map, $A_V({\bf x})$,
are defined as:
\begin{equation}
S_p(l)=\langle |A_V({\bf x})-A_V({\bf x}+{\bf l})|^p\rangle
\label{mom}
\end{equation}
where $p$ is the order and the average is extended to all map positions
${\bf x}$.
In turbulent flows it is found that the structure functions of velocity
are power laws. Assuming that the structure functions of projected density 
(or extinction) are power laws as well, we call $\eta(p)$ the exponents 
of these power laws:
\begin{equation}
S_p(l)\propto l^{\eta(p)}
\label{}
\end{equation}

In Figure~\ref{fig2} we have plotted the structure functions of the
map obtained with 10 stars per cell, relative to the third order 
structure function, since we are interested in investigating the
relative scaling, $\eta(p)/\eta(3)$ (this follows the idea of
extended self--similarity by Benzi et al. 1993 and Dubrulle 1994). 

\ifnum\astroph=1

\begin{figure}[h]
\epsfxsize=7cm \epsfbox{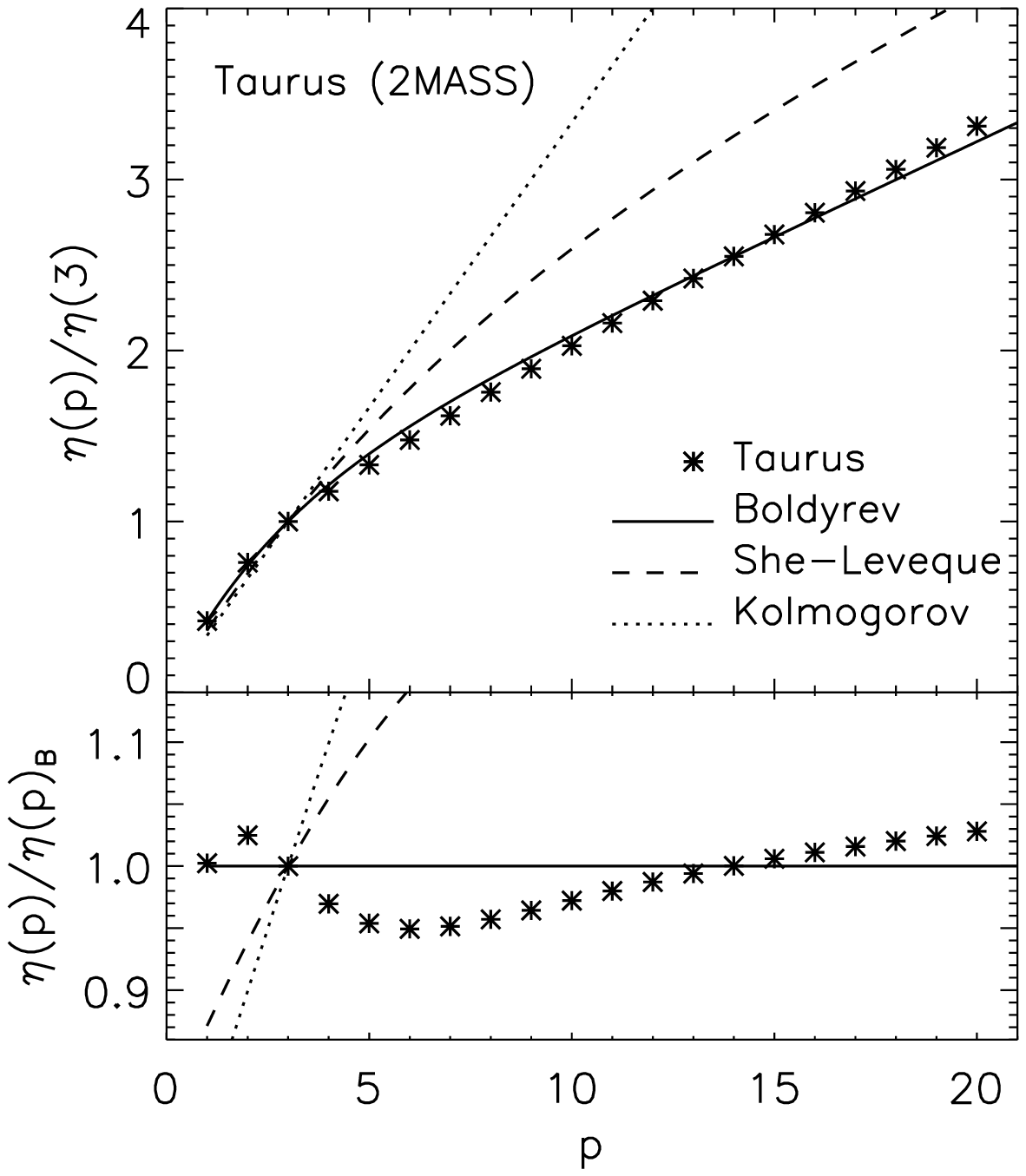}
\caption[]{Top panel: Relative structure function scaling. Bottom 
panel: Structure function exponents divided by the
Boldyrev's scaling.}
\label{fig3}
\end{figure}

\fi

The function $\eta(p)/\eta(3)$ is plotted in Figure~\ref{fig3}. 
The velocity scaling predicted by Boldyrev (2002) for supersonic 
turbulence is shown as a solid line, the one predicted by She \& 
Leveque (1994) for incompressible turbulence is plotted as a dashed line
and the Kolmogorov's velocity scaling, $p/3$, is shown by the dotted
line. The scaling of the structure functions of projected density
is found to be the same as the scaling of velocity structure functions
in supersonic turbulence.
{\it A priori} we do not know if the relative scaling of projected density
should be the same as that of the velocity; the velocity scaling from 
theoretical models is shown here only as a reference. 

\nocite{Boldyrev2002} \nocite{Benzi+93}
\nocite{Kolmogorov41} \nocite{She+Leveque94} \nocite{Dubrulle94}

Padoan et al. (2002) have recently analyzed in a similar way $^{13}$CO maps
of the same Taurus region and of Perseus. The present result for the 
relative scaling of the structure functions in Taurus confirms the 
results of that work. However, the sample size from the 2MASS data is
much larger than the sample size of the $^{13}$CO map. Furthermore, the
spatial resolution of the extinction map is slightly better 
and the range of column density sampled 20 times larger than in the
$^{13}$CO map. The statistical significance of high order moments
from the analysis of the 2MASS data should therefore be much higher 
than from the $^{13}$CO maps.

\nocite{Padoan+2002scaling}

\section{Conclusions}

Boldyrev (2002) has recently proposed an analytic model for the velocity 
structure function scaling in supersonic turbulence. The model is an 
extension of the scaling of incompressible turbulence proposed by
She \& Leveque (1994) and has already been successfully tested with 
numerical simulations of supersonic turbulence (Boldyrev, Nordlund \& 
Padoan 2002a). An equivalent analytic model for the scaling of the 
structure functions of projected density in supersonic turbulence is
not available yet. Only the slope of the second order structure function 
has been derived from the velocity structure functions, under certain 
approximations (Boldyrev, Nordlund \& Padoan 2002b). However, the fact
that the projected density follows the same scaling as the velocity 
field in supersonic turbulence suggests that the density field in the 
Taurus region is the result of a multiplicative process with a Log-Poisson
statistics (Dubrulle 1994), very likely the result of the turbulent
fragmentation. 

\nocite{Boldyrev+2002scaling} \nocite{Boldyrev+2002structure} 
\nocite{She+Leveque94} \nocite{Dubrulle94}

The importance of supersonic turbulence in the fragmentation of 
star-forming regions has been established in previous works (e.g. 
Padoan \& Nordlund 1999; Padoan et al. 2001; Padoan \& Nordlund 2002).
The purpose of the present work is primarily to determine
the statistical properties of the fragmentation process, independent of 
its origin. Such statistical properties may be universal, for example 
if they are mainly the consequence of turbulence, or depend on several 
physical parameters, such as gas density, temperature, turbulent 
velocity dispersion and star formation activity. We plan to compute 
and analyze 2MASS extinction maps of different extended star-forming 
regions in order to establish the properties of the fragmentation process 
that leads to the formation of stars in different environments. 

\nocite{Padoan+Nordlund99mhd} \nocite{Padoan+2001cores} 
\nocite{Padoan+Nordlund2002imf}

\acknowledgements

This publication makes use of data products from the Two Micron All Sky
Survey, which is a joint project of the
University of Massachusetts and the Infrared Processing and Analysis
Center/California Institute of Technology, funded
by the National Aeronautics and Space Administration and the National Science
Foundation.

The work of PP was performed while PP held a National Research 
Council Associateship Award at the Jet Propulsion Laboratory, 
California Institute of Technology.

\clearpage


\ifnum\astroph=0
\clearpage

\onecolumn

{\bf Table and Figure captions:} \\

{\bf Table \ref{t1}:} Extinction map parameters for different spatial 
resolutions. \\

{\bf Figure \ref{fig1}:} Extinction map of the Taurus region computed 
with 10 stars per cell. \\

{\bf Figure \ref{fig2}:} Structure functions of the extinction map from 
$p=1$ to 20 relative to the third order structure function. The extinction 
map with 10 stars per cell has been used. The solid lines show the least 
square fits used to define the power law slopes. \\

{\bf Figure \ref{fig3}:} Top panel: Relative structure function scaling. 
Bottom panel: Structure function exponents divided by the
Boldyrev's scaling. \\

\clearpage
\begin{table}
\begin{tabular}{cccc}
\hline
\hline
No. of stars & $\sigma_{A_V}$ & Median Resolution & $A_{V,{\rm max}}$  \\
             &  (mag)         & (arcmin)          & (mag) \\
\hline
100 & 0.26 & 12.4 & 8.0 \\
30  & 0.34 & 6.7  & 11.7 \\
10  & 0.49 & 3.4  & 19.5 \\
3   & 0.85 & 1.7  & 26.3 \\
1   & 1.30 & 1.0  & 32.7 \\
\hline
\end{tabular}
\caption{Extinction map parameters for different spatial resolutions}
\label{t1}
\end{table}

\clearpage
\begin{figure}
\centerline{
\epsfxsize=13cm \epsfbox{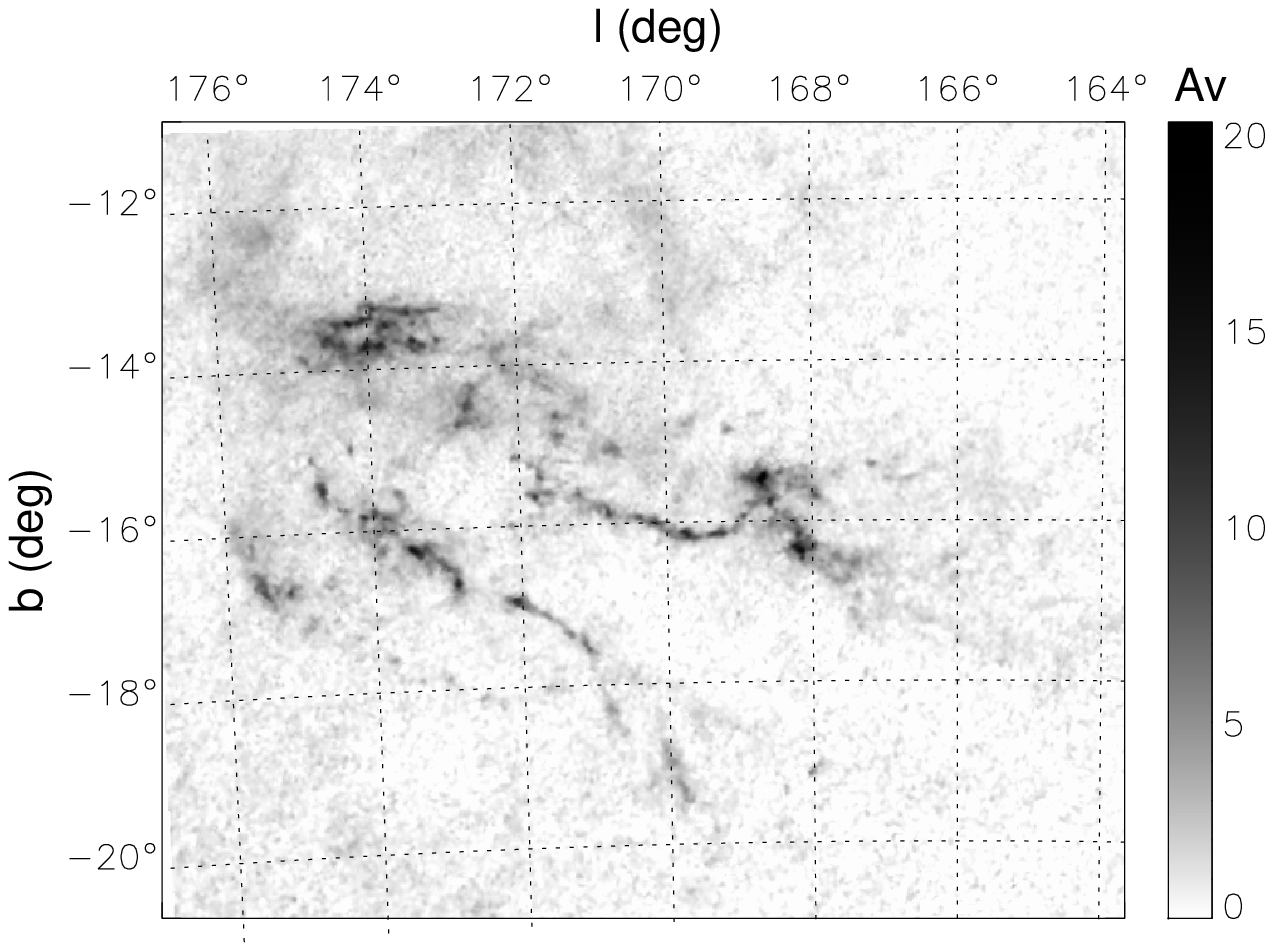}
}
\caption[]{}
\label{fig1}
\end{figure}

\clearpage
\begin{figure}
\centerline{
\epsfxsize=13cm \epsfbox{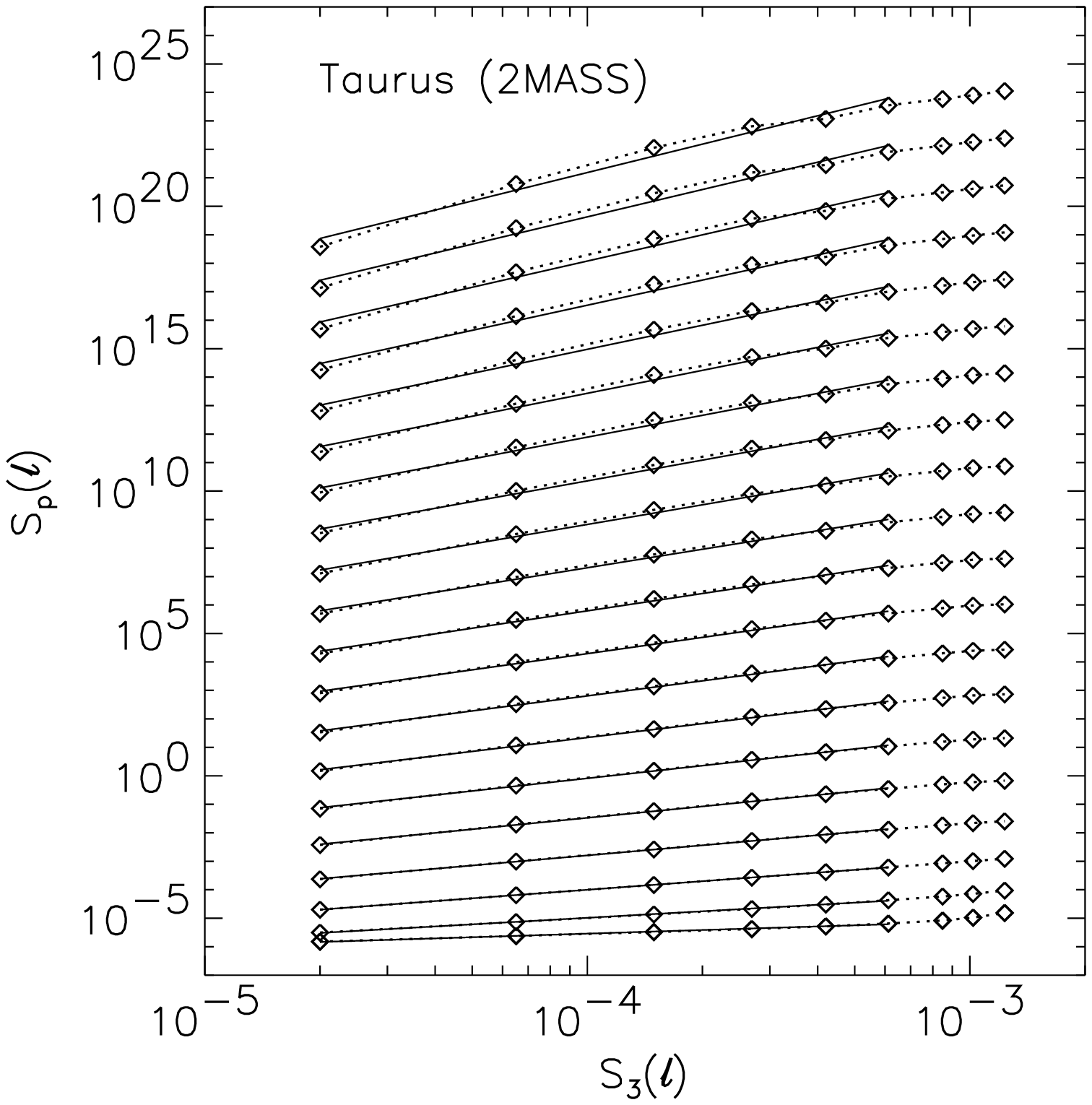}
}
\caption[]{}
\label{fig2}
\end{figure}

\clearpage
\begin{figure}
\centerline{
\epsfxsize=10cm \epsfbox{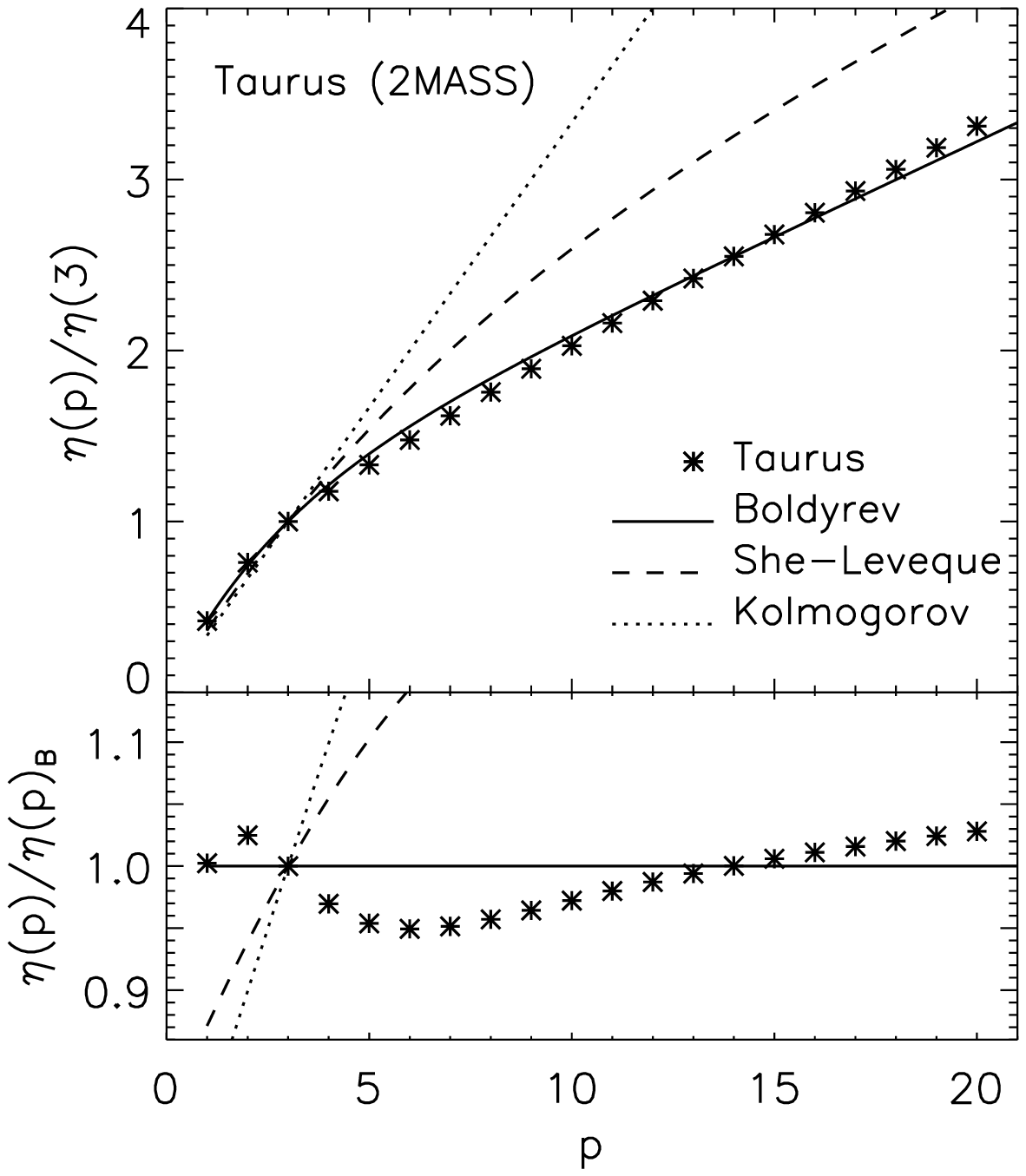}
}
\caption[]{}
\label{fig3}
\end{figure}

\fi

\end{document}